\documentclass[a4paper]{spie}  
\usepackage[]{graphicx}

\newcommand{\LAOG}{a}
\newcommand{\Cavendish}{b}
\newcommand{\MPIfR}{c}
\newcommand{\CAUP}{d}
\newcommand{\INAFOATo}{e}
\newcommand{\INAFOAR}{f}
\newcommand{\IAGL}{g}
\newcommand{\IfAW}{h}
\newcommand{\AIU}{i}
\newcommand{\CEALETI}{j}
\newcommand{\CRAL}{k}
\newcommand{\FCUL}{l}
\newcommand{\INETI}{m}
\newcommand{\INAFOAA}{n}
\newcommand{\JMMC}{o}
\newcommand{\ESOChile}{p}
\newcommand{\ESOHQ}{q}
\newcommand{\IAC}{r}
\newcommand{\OCA}{s}
\newcommand{\GRAAL}{t}

\title{VSI: the VLTI spectro-imager}

\author{%
F. Malbet \supit{\LAOG} 
D. Buscher\supit{\Cavendish}
G. Weigelt\supit{\MPIfR}
P. Garcia\supit{\CAUP}
M. Gai\supit{\INAFOATo}
D. Lorenzetti\supit{\INAFOAR}
J. Surdej\supit{\IAGL}
J. Hron\supit{\IfAW}
R. Neuh\"auser\supit{\AIU}
P. Kern\supit{\LAOG}
L. Jocou\supit{\LAOG}
J.-P. Berger\supit{\LAOG}
O. Absil\supit{\LAOG}
U. Beckmann\supit{\MPIfR}
L. Corcione\supit{\INAFOATo}
G. Duvert\supit{\LAOG,\JMMC}
M. Filho\supit{\CAUP}
P. Labeye\supit{\CEALETI}
E. Le Coarer\supit{\LAOG}
G. Li Causi\supit{\INAFOAR}
J. Lima\supit{\FCUL}
K. Perraut\supit{\LAOG}
E. Tatulli\supit{\LAOG,\INAFOAA,\JMMC}
E. Thi\'ebaut\supit{\CRAL}
J. Young\supit{\Cavendish}
G. Zins\supit{\LAOG}
A. Amorim\supit{\FCUL}
B. Aringer\supit{\IfAW}
T. Beckert\supit{\MPIfR}
M. Benisty\supit{\LAOG}
X. Bonfils\supit{\FCUL}
A. Cabral\supit{\INETI}
A. Chelli\supit{\LAOG,\JMMC}
O. Chesneau\supit{\OCA}
A. Chiavassa\supit{\GRAAL}
R. Corradi\supit{\IAC}
M. de Becker\supit{\IAGL}
A. Delboulb\'e\supit{\LAOG}
G. Duch\^ene\supit{\LAOG}
T. Forveille\supit{\LAOG}
C. Haniff\supit{\Cavendish}
E. Herwats\supit{\LAOG,\IAGL}
K.-H.Hofmann\supit{\MPIfR}
J.-B. Le Bouquin\supit{\ESOChile}
S. Ligori\supit{\INAFOATo}
D. Loreggia\supit{\INAFOAR}
A. Marconi\supit{\INAFOAA}
A. Moitinho\supit{\FCUL}
B. Nisini\supit{\INAFOAR}
P.-O. Petrucci\supit{\LAOG}
J. Rebordao\supit{\INETI}
R. Speziali\supit{\INAFOAR}
L. Testi\supit{\INAFOAA,\ESOHQ}
F. Vitali\supit{\INAFOAR}
\skiplinehalf
\supit{\LAOG}
Universit\'e J.~Fourier, CNRS, Laboratoire d'Astrophysique de Grenoble,
UMR 5571, BP 53, F-38041 Grenoble cedex 9, France; \\
\supit{\Cavendish}
Cavendish Laboratory of University of Cambridge, UK; \\
\supit{\MPIfR}
Max-Planck Institute for Radioastronomy, Bonn, Germany; \\
\supit{\CAUP}
Faculdade de Engenharia \& Centro de Astrofísica, Universidade do Porto, Portugal\\
\supit{\INAFOATo}
INAF - Osservatorio Astronomico di Torino, Italy; \\
\supit{\INAFOAR}
INAF - Osservatorio Astronomico di Roma, Italy; \\
\supit{\IAGL}
Institute of Astrophysics and Geophysics,  Li\`ege, Belgium; \\
\supit{\IfAW}
Institute of Astrophysics of the university of Wien, Austria; \\
\supit{\AIU}
Astrophysical Institute and University Observatory, Jena, Germany; \\
\supit{\CEALETI}
CEA-LETI, Minatec, Grenoble, France; \\
\supit{\CRAL}
Centre de Recherche en Astrophysique de Lyon, France; \\
\supit{\FCUL}
SIM/IDL Faculdade de Ci\^encias da Universidade de Lisboa, Portugal; \\
\supit{\INETI}
Instituto Nacional de Engenharia, Tecnologia e Inovacco, Lisboa, Portugal; \\
\supit{\INAFOAA}
INAF/Osservatorio di Astrofisica di Arcetri, Italy; \\
\supit{\JMMC}
Jean-Marie Mariotti Center, CNRS, France; \\
\supit{\ESOChile}
European Southern Observatory, Santiago, Chile; \\
\supit{\ESOHQ}
European Southern Observatory Headquarters, Garching, Germany; \\
\supit{\IAC}
Instituto de Astrof\'isica de Canarias, Spain; \\
\supit{\OCA}
Observatoire de la C\^ote d'Azur, Laboratoire Gemini, Nice, France; \\
\supit{\GRAAL}
Groupe de Recherches en Astronomie et Astrophysique du Languedoc, Montpellier, France
}

\authorinfo{Send correspondence to F.~Malbet: \texttt{Fabien.Malbet@obs.ujf-grenoble.fr}}

\begin{document} 
\maketitle 

\begin{abstract}
  The VLTI Spectro Imager (VSI) was proposed as a second-generation
  instrument of the \emph{Very Large Telescope Interferometer}
  providing the ESO community with spectrally-resolved, near-infrared
  images at angular resolutions down to 1.1 milliarcsecond and
  spectral resolutions up to $R=12000$. Targets as faint as $K=13$
  will be imaged without requiring a brighter nearby reference object;
  fainter targets can be accessed if a suitable reference is
  available. The unique combination of high-dynamic-range imaging at
  high angular resolution and high spectral resolution enables a
  scientific program which serves a broad user community and at
  the same time provides the opportunity for breakthroughs in many
  areas of astrophysics.  The high level specifications of the
  instrument are derived from a detailed science case based on the
  capability to obtain, for the first time, milliarcsecond-resolution
  images of a wide range of targets including: probing the initial
  conditions for planet formation in the AU-scale environments of
  young stars; imaging convective cells and other phenomena on the
  surfaces of stars; mapping the chemical and physical environments of
  evolved stars, stellar remnants, and stellar winds; and
  disentangling the central regions of active galactic nuclei and
  supermassive black holes. VSI will provide these new capabilities
  using technologies which have been extensively tested in the past
  and VSI requires little in terms of new infrastructure on the
  VLTI. At the same time, VSI will be able to make maximum use of new
  infrastructure as it becomes available; for example, by combining 4,
  6 and eventually 8 telescopes, enabling rapid imaging through
  the measurement of up to 28 visibilities in every wavelength channel
  within a few minutes. The current studies are focused on a
  4-telescope version with an upgrade to a 6-telescope one. The
  instrument contains its own fringe tracker and tip-tilt control in
  order to reduce the constraints on the VLTI infrastructure and
  maximize the scientific return.
\end{abstract}



\section{Introduction}
\label{sec:introduction}

At the beginning of the 21st century, infrared observations performed
at the milli-arcsecond scale are essential for many astrophysical
investigations either to compare the same physical phenomena at
different wavelengths (like sources already observed with the VLBI or
soon to be observed by ALMA) or to get finer details on observations
carried out with the \emph{Hubble Space Telescope} (HST) or 10-m class
telescopes equipped with adaptive optics.  The astrophysical science
cases at milli-arcsecond scales which range from planetary physics to
extragalactic studies can only be studied using interferometric
aperture synthesis imaging with several optical telescopes.  In this
respect, the \emph{Very Large Telescope} (VLT) observatory of the
\emph{European Southern Observatory} (ESO) is a unique site world-wide
with $4\times8$-m unit telescopes (UTs), $4\times1.8$-m auxiliary
telescopes (ATs) and all the required infrastructure, in particular
delay lines (DLs), to combine up to 6 telescopes. The \emph{VLT
  Interferometer} (VLTI) infrastructure can be directly compared to
the \emph{Plateau de Bure Interferometer} (PdBI) which combines
$6\times15$-m antenna over 500-m in the millimeter-wave domain. The
quality of the foreseen images can be directly compared to the images
provided by the PdBI. However, the angular resolution of the VLTI is a
few hundred times higher due to the observation at shorter
wavelengths.  The large apertures of the VLTI telescopes and the
availability of fringe tracking allow sensitivity and spectral
resolution to be added to the imaging capability of the VLTI.

In April 2005, at the ESO workshop on \emph{``The power of
  optical/infrared interferometry: recent scientific results and
  second generation VLTI instrumentation''}, two independent teams
have proposed two different concepts for an imaging near-infrared
instrument for the VLTI: BOBCAT \cite{2008poii.conf..407B} and VITRUV
\cite{2008poii.conf..357M}. In October 2005, the science cases of
these instruments were approved by the ESO \emph{Science and Technical
  Committee}. In January 2006, the two projects merged in order to
propose the \emph{VLTI spectro-imager} (VSI) as a response to the ESO
call for phase A proposals for second generation VLTI instruments. The
phase A study ended in September 2007 after an ESO board review. The
VSI instrument has been recommended by the ESO \emph{Science and
  Technical Committee} in October 2007 and ESO is currently proposing
to proceed with the construction in 2009-2010 for an installation of
the 4/6 telescope version at the VLT in 2015. In this volume, we
present the result of the VSI phase A study.

\section{VSI overview}
\label{sect:overview}

The VLTI Spectro Imager will provide the astronomical community with
spectrally-resolved near-infrared images at angular resolutions down
to 1.1 milliarcsecond and spectral resolutions up to $R=12000$.
Targets as faint as $K=13$ will be imaged without requiring a brighter
nearby reference object; fainter targets can be accessed if a suitable
off-axis reference is available. This unique combination of
high-dynamic-range imaging at high angular resolution and high
spectral resolution for a wide range of targets enables a scientific
program which will serve a broad user community and at the same time
provide the opportunity for breakthroughs in many areas at the
forefront of astrophysics.

A great advantage of VSI is that it will provide these new
capabilities while using techniques which have extensively been
experimented in the past. VSI will be capable of making maximum use of
the new VLTI infrastructure as it becomes available.  Operations with
less than 8 telescopes are the scope of the first phases of VSI.  Three
development phases are foreseen: VSI4 combining 4 telescopes (UTs or
ATs), VSI6 combining 6 telescopes (4UTS+2ATs or 4ATs+2UTS and
eventually 6ATs), and perhaps in the long-run, VSI8
combining 8 telescopes (4UTs+4ATs or eventually 8ATs). The current
study is focused on a 4-telescope version with an upgrade to a
6-telescope one. The instrument contains its own fringe tracker and
tip-tilt control in order to reduce the constraints on the VLTI
infrastructure and maximize the scientific return.

\section{Science cases for VSI}
\label{sect:science}

The high level specifications of the instrument are derived from
science cases based on the capability to reconstruct
milli-arcsecond-resolution images of a wide range of
targets. These science cases are detailed by Garcia et al.\cite{SPIE-7013-172} in this
volume, but we list here the 4 main scientific cases:
\begin{itemize}
\item \textbf{Formation of stars and planets}. The early evolution of
  stars and the initial conditions for planet formation are determined
  by the interplay between accretion and outflow processes. Due to the
  small spatial scales where these processes take place,
  very little is known about the actual physical and chemical
  mechanisms at work.  Interferometric imaging at 1 milli-arcsecond
  spatial resolution will directly probe the regions responsible for
  the bulk of excess continuum emission from these objects, therefore
  constraining the currently highly degenerate models for the spectral
  energy distribution (see Fig.~\ref{fig:yso}).
  \begin{figure}[t]
    \centering
    \includegraphics[width=0.95\hsize]{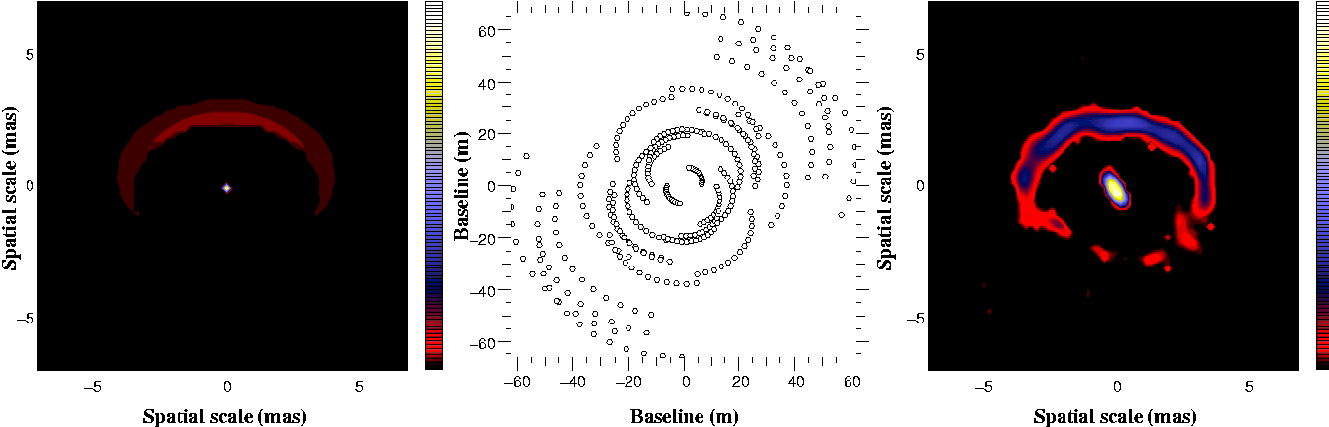}
    \caption{Image reconstruction performed with 6 ATs on a model disk
      around an Herbig Ae star. Left: model image;
      middle: coverage of the spatial frequencies; right:
      reconstructed image. The dust structure, the inner dust radius
      and the asymmetry (vertical structure) are well retrieved.
      Relative photometry is reliable (17\% vs 19\% of flux in the
      central star).}
    \label{fig:yso}
  \end{figure}
  In the emission lines a variety of processes will be probed, in
  particular outflow and accretion magnetospheres.  The inner few AUs
  of planetary systems will also be studied, providing additional
  information on their formation and evolution processes, as well as
  on the physics of extrasolar planets. Renard et
  al.\cite{SPIE-7013-107} in this volume detail the case of the
  detection of extrasolar planets with VSI.
\item \textbf{Imaging stellar surfaces}. Optical and near-infrared
  imaging instruments provide a powerful means to resolve stellar features
  of the generally patchy surfaces of stars throughout the
  Hertzsprung-Russell diagram.  Optical/infrared interferometry has
  already proved its ability to derive surface structure parameters
  such as limb darkening or other atmospheric parameters.  VSI, as an
  imaging device, is of strong interest to study various specific
  features such as vertical and horizontal temperature profiles and
  abundance inhomogeneities, and to detect their variability as the
  star rotates.  This will provide important keys to address stellar
  activity processes, mass-loss events, magneto-hydrodynamic
  mechanisms, pulsation and stellar evolution.
\item \textbf{Evolved stars, stellar remnants \& stellar winds}. HST
  and ground-based observations revealed that the geometry of young
  and evolved planetary nebulae and related objects (e.g., nebulae
  around symbiotic stars) show an incredible variety of elliptical,
  bi-polar, multi-polar, point-symmetrical, and highly collimated
  (including jets) structures. The proposed mechanisms explaining the
  observed geometries (disks, magnetohydrodynamics collimation and
  binarity) are within the grasp of interferometric imaging at 1~mas
  resolution.  Extreme cases of evolved stars are stellar black holes.
  In microquasars, the stellar black-hole accretes mass from a donor.
  The interest of these systems lies in the small spatial scales and
  high multi-wavelength variability. Milliarcsecond imaging in the
  near-infrared will allow disentangling between dust and jet synchrotron
  emission, comparison of the observed morphology with radio maps and
  correlation of the morphology with the variable X-ray spectral
  states.
\item \textbf{Active Galactic Nuclei \& Supermassive Black Holes}. AGN
  consist of complex systems composed of different interacting parts powered
  by accretion onto the central supermassive black hole. The imaging
  capability will permit the study of the geometry and dust composition of
  the obscuring torus and the testing of radiative transfer models (see
  Fig.~\ref{fig:agn}).
  \begin{figure}[t]
    \centering
    \includegraphics[width=0.95\hsize]{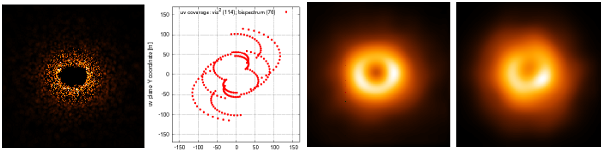}
    \caption{VLTI/VSI image reconstruction simulation performed with 4
      UTs on a model of a clumpy torus at the center of an AGN. Left:
      model torus image; middle left: simulated coverage of the
      spatial frequencies corresponding to the southern AGN NGC~1365
      at declination of $-36^{\circ}$; middle right: model image
      convolved with a perfect beam corresponding to the maximum
      spatial resolution; right: image reconstructed from simulated
      VSI data using the Building Block method. }
  \label{fig:agn}
  \end{figure}
  Milli-arcsecond resolution imaging will allow us to probe the
  collimation at the base of the jet and the energy distribution of
  the emitted radiation.  Supermassive black hole masses in nearby
  (active) galaxies can be measured and it will be possible
  to detect general relativistic effects for the stellar orbits closer
  to the galactic center black hole. The wavelength-dependent
  differential-phase variation of broad emission lines will provide
  strong constraints on the size and geometry of the Broad Line Region
  (BLR).  It will then be possible to establish a secure
  size-luminosity relation for the BLR, a fundamental ingredient to
  measure supermassive black hole masses at high redshift.
\end{itemize}

We have shown that this astrophysical program could provide the
premises for a legacy program at the VLTI. For this goal, the number
of telescopes to be combined should be at least 4, or better 6 to 8 at
the VLTI at the time when the \emph{James Webb Space Telescope} will
hammer faint infrared science ($\sim2013$), when HAWK-I, KMOS, will
have hopefully delivered most of their science, and ALMA will be fully
operational.  The competitiveness and uniqueness of the VLT will
remain on the high angular (AO/VLTI) and the high spectral resolution
domains. In a context where the European \emph{Extremely Large
  Telescope} (ELT) will start being constructed, then have first
light, and, where Paranal science operations will probably be
simplified with less VLT instruments and an emphasis on survey
programs, VSI will take all its meaning by bringing the VLTI to a
legacy mode.

\section{Astrophysical specifications}
\label{sec:astr-spec}

\begin{table}[t]
  \centering
  \caption{VSI specifications}
  \label{tab:specs}
  \smallskip
  \begin{tabular}{|ll|}
    \hline
    Criteria &Specifications \\
    \hline
    Time resolution &Minimum: 1 night\\
    Data set interval &30 min \\                                       
    Spectral coverage &$J,H,K$\\
    Exposure time	&10 ms - 100s \\                                    
    Spectral resolution &100, 2000, (5000), 12000\\
    Polarization	&No scientific requirements \\                       
    Faintest objects &$J$: 12, $H$: 12, $K$: 15\\
    Science data product	&Raw data with OI-fits files \\             
    Dynamic range &1/100 (Standard), 1/1000 (High dynamic), 1/10000 (Parametric)\\
    Environmental data &Turbulence parameters, AO/Tip-Tilt and Fringe Tracker performance \\ 
    Image complexity &$10 \times 10$ pixels\\
    Image reconstruction	 &Image reconstruction software available \\   
    Field of View &$J$-UT: 30 mas to $K$-AT: 250 mas\\
    Instrument life expectancy	 &10 years  \\                      
    \hline
  \end{tabular}
\end{table}

The astrophysical and technical specifications of VSI are listed in
Table \ref{tab:specs}. The three first lines of this table are the
ones that makes VSI unique.

As a matter of fact in the VSI science case, which is the study at
high spatial and spectral resolutions of compact astrophysical
sources, the variation of the morphology of the objects can be quite
rapid. For example the RS Ophiuci\cite{2007A&A...464..119C} novae
expansion was measured 5 days after outburst by AMBER and 14 days
after outburst in radio. The sizes doubled from about 5\,mas to about
10\,mas in less than 10 days. As another example, at Taurus distance,
Keplerian period at 1 mas from the star is 20 days. When magnifying at
higher angular resolution, it is expected that the phenomena at stake
vary more rapidly. Our understanding is that VSI should be able to
image within a night and therefore avoid any changes of array
configurations.

\begin{figure}[p]
  \centering
  \includegraphics[width=\textwidth]{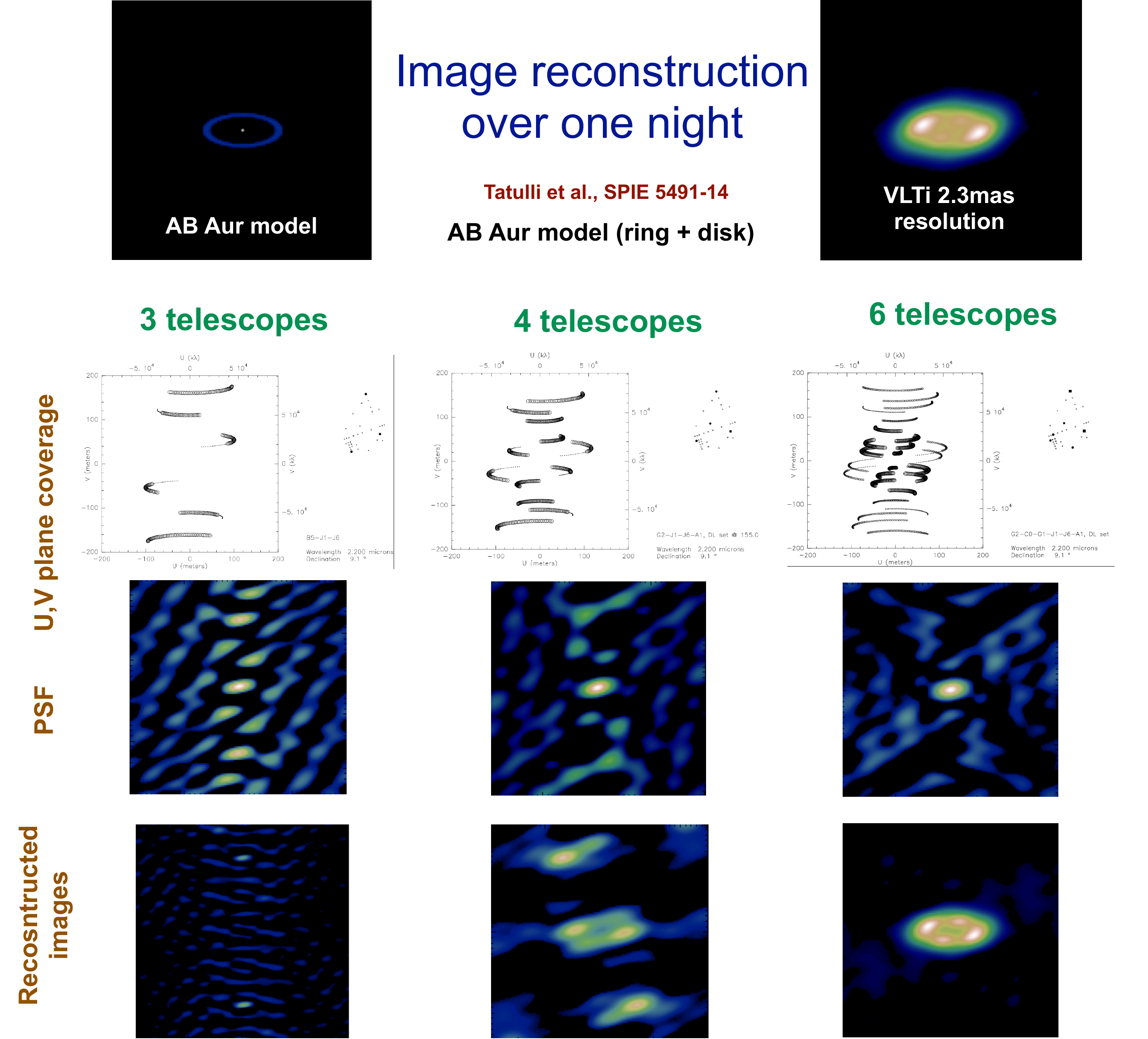}
  \caption{Simulated image for the young star AB Aurigae reconstructed
    with 3 different VLTI configuration in one night: left with 3
    telescopes, middel with 4 telescopes and right with 6
    telescopes. Top left is the original model and top right is the
    model has would be seen by a single telescope of the size of the
    VLTI. From Tatulli\cite{2004SPIE.5491..117T} et al. 2004.}
  \label{fig:abaur}
\end{figure}
Therefore, it is important to understand how many telescopes are
required to obtain an image within a night. Figure~\ref{fig:abaur}
illustrates this argument in the case of a disk around a young
star. One sees that 6 telescopes are required to get an sensible image
within a night\cite{2004SPIE.5491..117T}. In fact to obtain the same
quality would require 3 nights with different configuration of the
VLTI with 4 telescopes and 7 nights with 3 telescopes, like in the
case of AMBER.

Another important specification of VSI is the spectral
resolution. With spectral resolution, one is able to identify tracers
of different species like for the gas atomic lines (Br$\gamma$,
carbon, oxygen) or molecules (H$_2$, CO, ...) or for dust the
continuum or silicate lines. Spectral resolution allows also the
observed regions to be probe kinematically: for example distinguish
between solid and differential rotation for stars, identify Keplerian
motion in disks, detect the wind kine-morphology or measure expansion
(novae, SN,...). Finally spectral coverage over several astronomical
bands enables to probe the physical conditions of the environments by
measuring color temperature. It allows also $(u,v)$ planes to be
extended by wavelength super-synthesis.

Other important but more classical specifications are the limiting
magnitudes required to detect the faintest objects of the different
categories, the dynamic range expected in the image and the complexity
of these images (see Table~\ref{tab:specs}). Several studies on image
reconstruction can be found elsewhere in this
volume.\cite{SPIE-7013-50,SPIE-7013-146}.

\section{Instrument concept}

\begin{figure}[p]
  \centering
  \includegraphics[width=0.5\hsize]{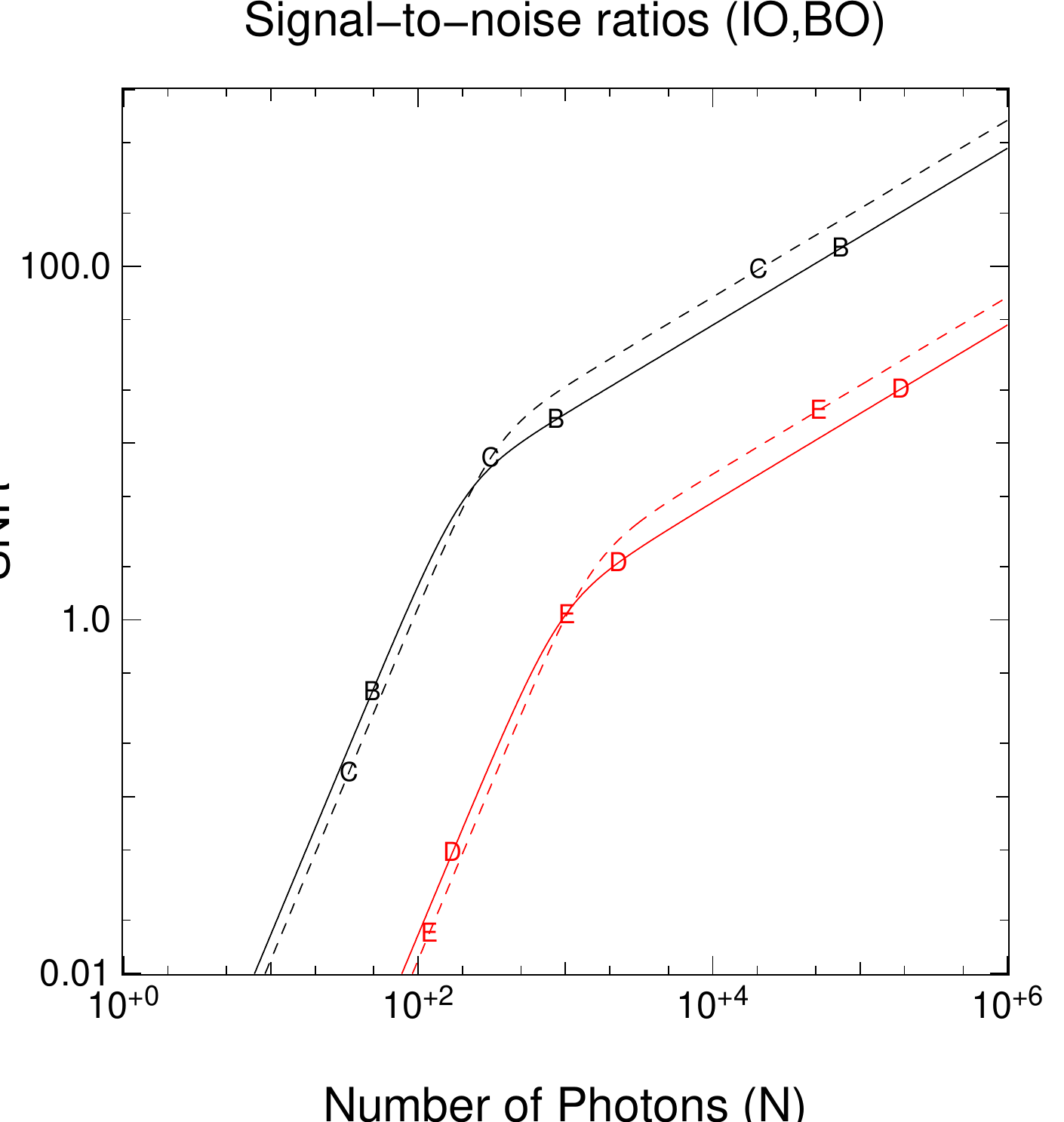}
  \caption{Signal-to-noise ratios computed in $H$-band for the IO
    combiner (solid lines) and the BO combiner (dashed lines) with a
    respective transmission of 65\% and 90\%. The black lines
    corresponds to a visibility of $1$ and the red lines to a
    visibility of $0.1$. The background noise is ignored.}
  \label{fig:snrs}
\end{figure}
\begin{figure}[p]
  \centering
  \includegraphics[width=0.5\hsize]{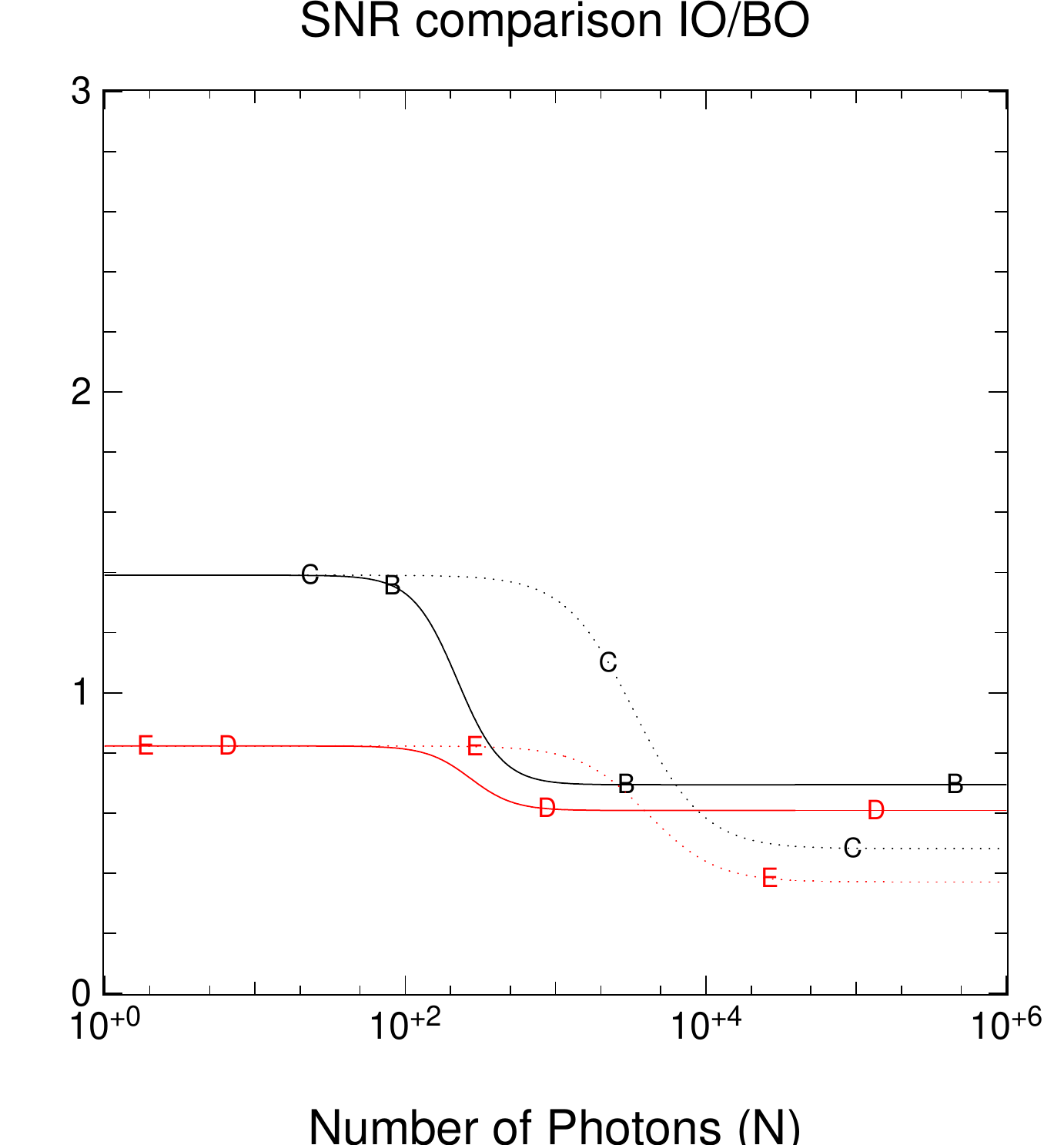}
  \caption{Relative comparison between the signal-to-noise ratio in
    $H$-band (black) and $K$-band (red) for the IO combiner and the BO
    combiner with a visibility of 1 (solid line) in same conditions as
    in Fig.~\ref{fig:snrs}. The dashed line corresponds to the limit
    where the visibility tends toward zero.}
  \label{fig:snriobo}
\end{figure}

The phase A study has led to an instrument concept which is presented
elsewhere in this volume\cite{SPIE-7013-106} consisting of:
\begin{itemize}
\item Integrated optics multi-way beam combiners\cite{SPIE-7013-40} providing
  high-stability visibility and closure-phase measurements on multiple
  baselines;
\item A cooled spectrograph\cite{SPIE-7013-108} providing resolutions between $R=100$
  and $R=12000$ over the $J$, $H$, or $K$ bands;
\item An integrated high-sensitivity switchable H/K fringe tracker capable of
  real-time cophasing or coherencing of the beams from faint or resolved
  sources;
\item Hardware and software to enable the
  instrument to be aligned, calibrated and operated with minimum
  staff overhead.
\end{itemize}

These features act in synergy to provide a scientific capability which
is a step beyond existing instruments. Compared to the single closure
phase measured by AMBER, the 3 independent closure phases available by
VSI4, the 10 independent closure phases measured by VSI6 and the 21
independent closure phases measured by VSI8 will make true
interferometric imaging, a routine process at the VLTI.  The
capability to cophase on targets up to $K=10$ will allow long
integrations at high spectral resolutions for large classes of
previously inaccessible targets, and the capability to do
self-referenced coherencing on objects as faint as $K=13$. It will
allow imaging of targets for which no bright reference is
sufficiently close by. VSI will be able to provide spectrally and
spatially-resolved ``image cubes'' for an unprecedented number of
targets at unprecedented resolutions.

\begin{figure*}[p]
  \centering
  \includegraphics[width=\hsize]{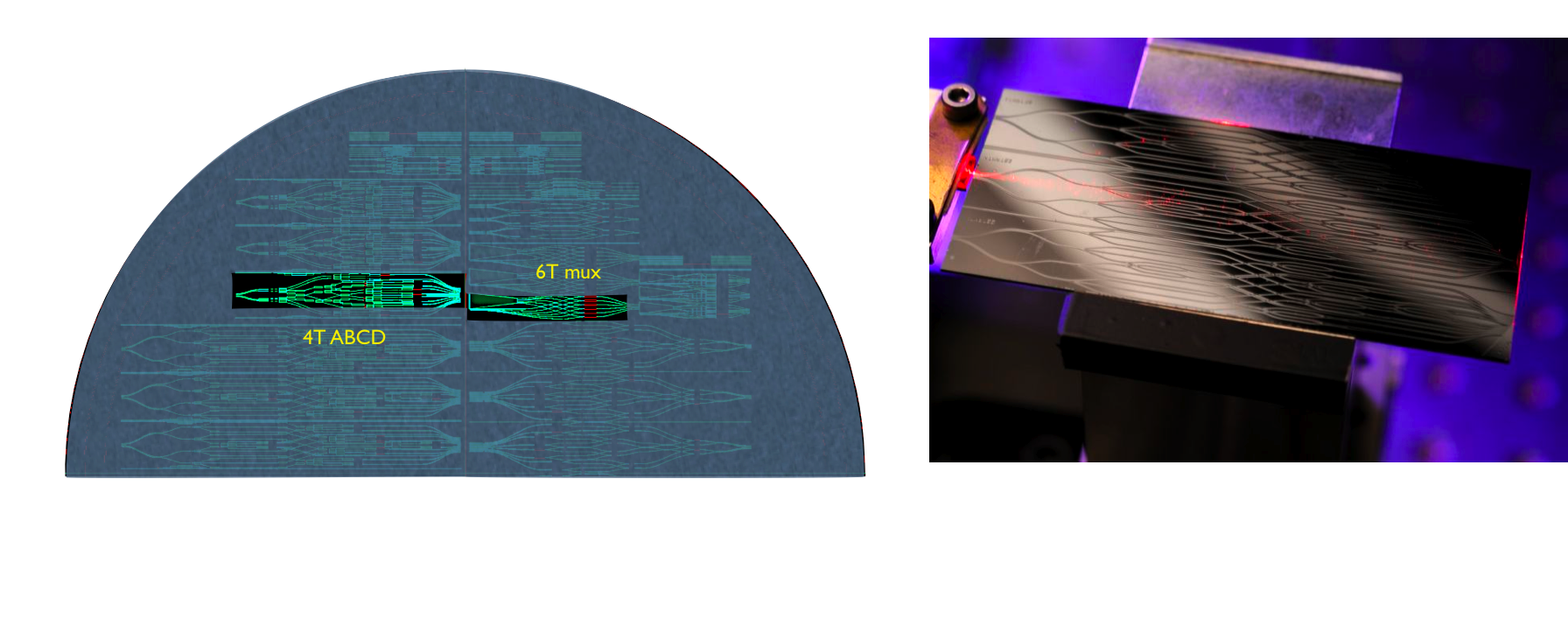}
  \caption{Science beam combiners\cite{SPIE-7013-40} of the VSI
    instrument: left the wafer with the 4T and 6T combiners enlighted
    and right the 4T component.}
  \label{fig:IOcomponents}
\end{figure*}
\begin{figure*}[p]
  \centering
  \includegraphics[width=0.8\hsize]{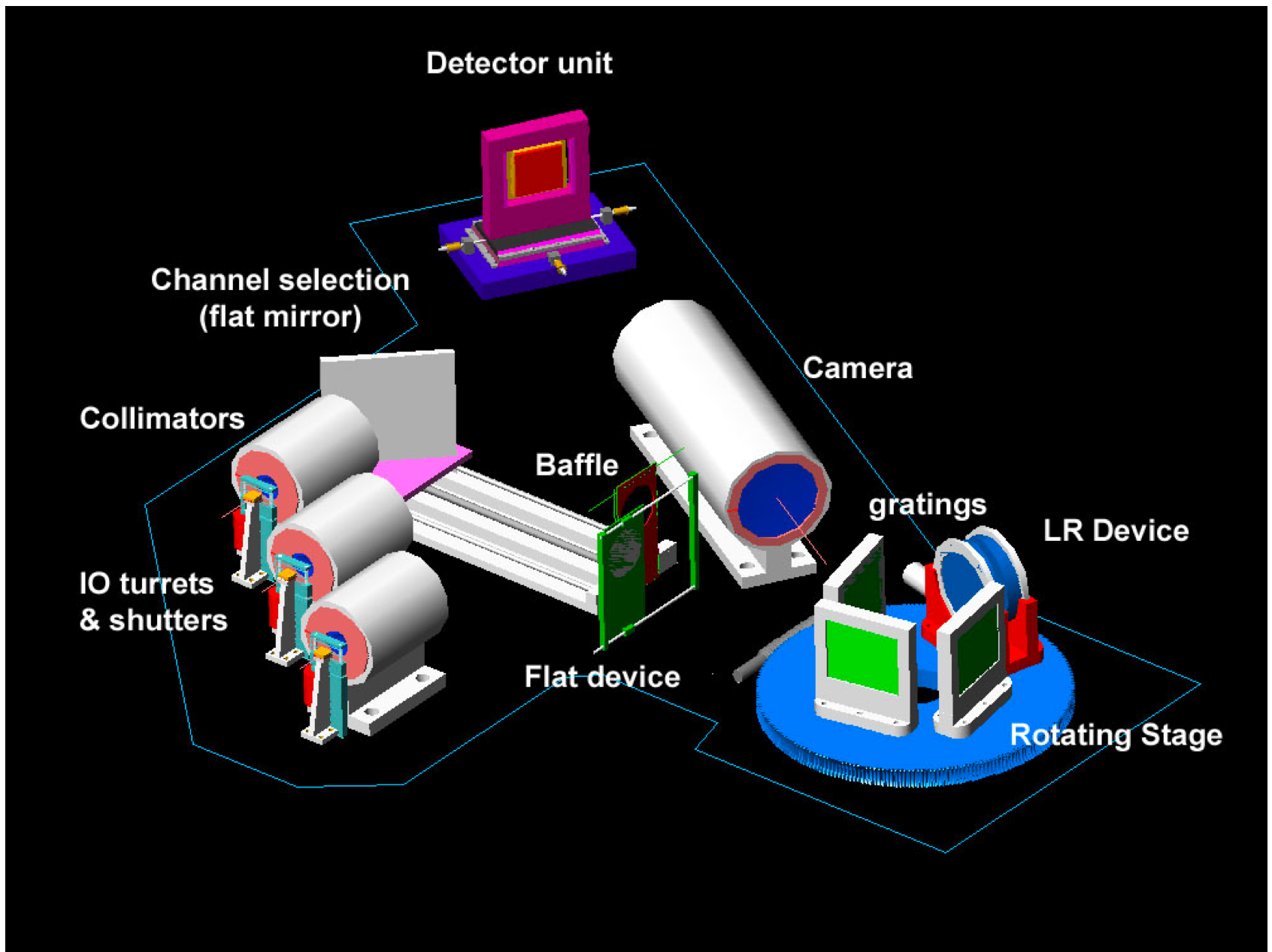}
  \caption{Implementation of the science
    instrument\cite{SPIE-7013-108} within the cryostat of the VSI
    instrument}
  \label{fig:SI}
\end{figure*}
\begin{figure*}[t]
  \centering
  \includegraphics[width=0.8\hsize]{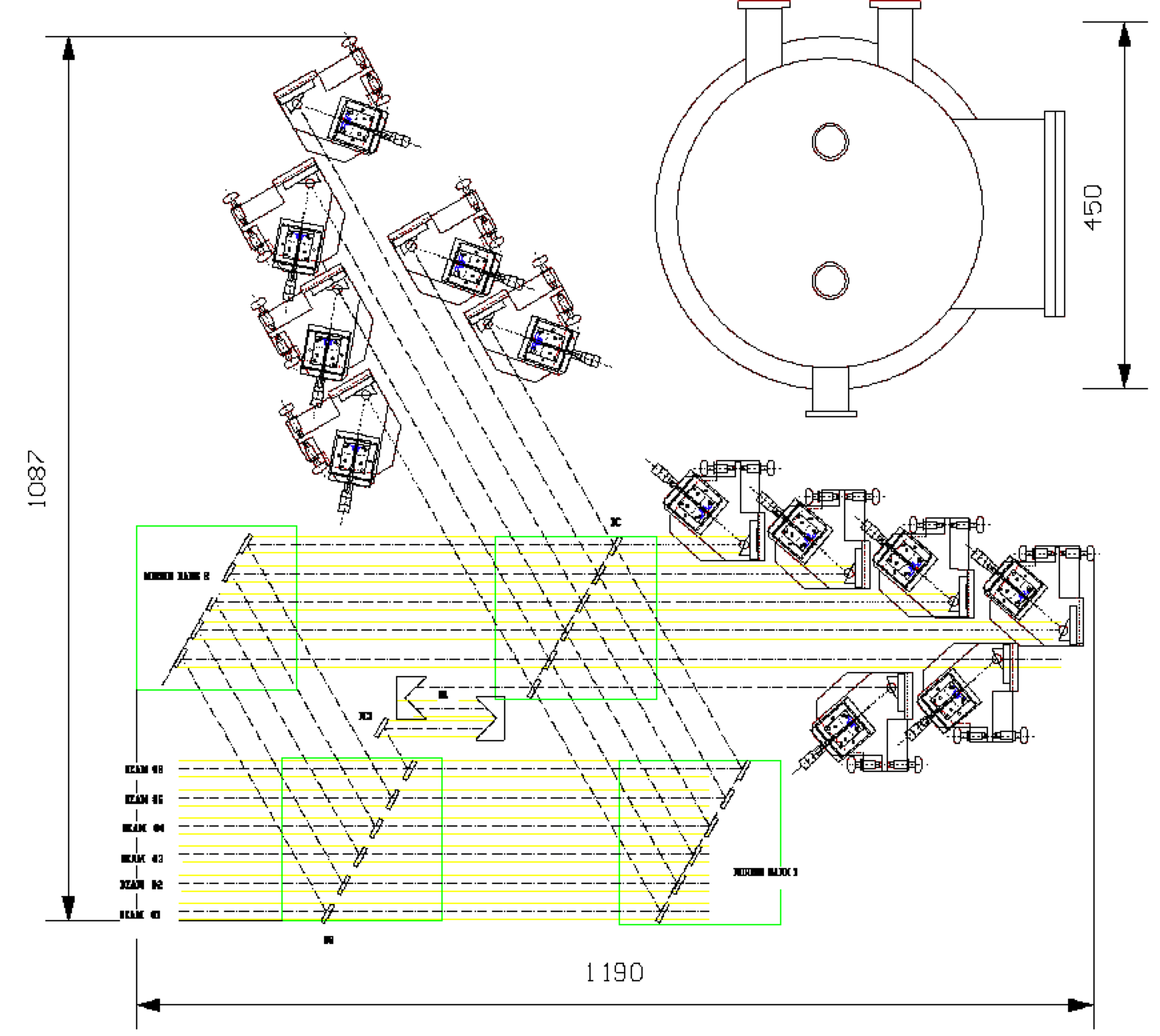}
  \caption{Fringe tracker concept\cite{SPIE-7013-152} of the VSI instrument}
  \label{fig:FT}
\end{figure*}
\begin{figure*}[t]
  \centering
  \includegraphics[width=\hsize]{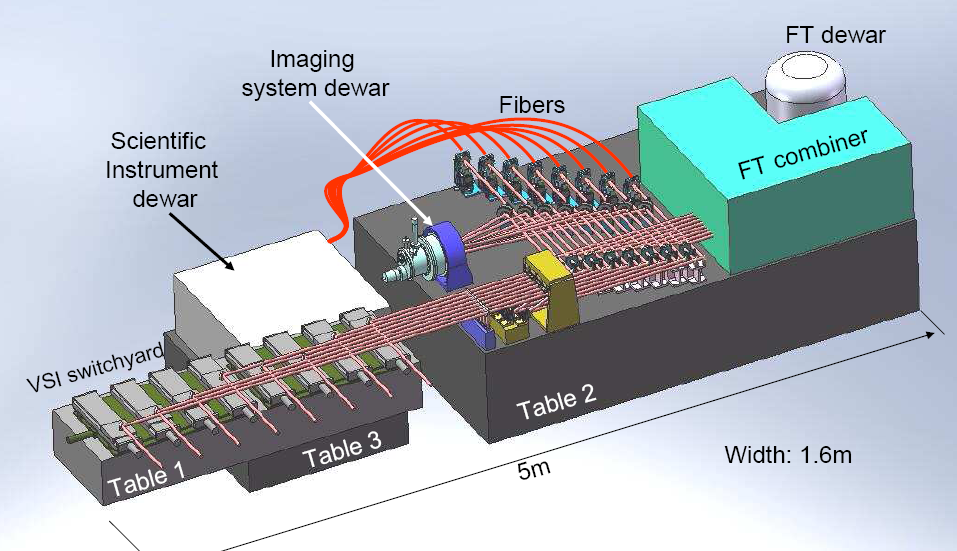}
  \caption{General implementation of the VSI instrument}
  \label{fig:implementation}
\end{figure*}
A system analysis of VSI has allowed the high level specifications of
the system to be defined, the external constraints to be clarified and
the functional analysis to be performed.  The system design
\cite{SPIE-7013-106} features 4 main assemblies: the science
instrument\cite{SPIE-7013-108} (SI, see Fig.~\ref{fig:SI}), the fringe
tracker\cite{SPIE-7013-152} (FT, see Fig.~\ref{fig:FT}), the common path (CP) and the
calibration and alignment tools (CAT). The global implementation is
presented in Fig.\ \ref{fig:implementation}.

The optics design of the science instrument features beam combination
using single mode fibers, an integrated optics chip\cite{SPIE-7013-40}
and 4 spectral resolutions through a cooled spectrograph described by
Lorenzetti et al.\cite{SPIE-7013-108} in this volume. We showed that
beam combiners in integrated optics or in bulk optics are comparable
in terms of SNR performances by comparing the beam combiners of BOBCAT
and VITRUV. Figures \ref{fig:snrs} and \ref{fig:snriobo} show that IO is
better in the photon-starved regime whereas the BO is better in the
photon-rich regime, but the gain is all cases limited to 30\%. The
integrated optics solution has been chosen for additional reasons like
maintainability, easiness of operation and availability of manpower.
The common path includes low-order adaptive optics (with the current
knowledge reduced to only tip-tilt corrections).  VSI also features an
internal fringe tracker (see \ref{fig:FT}) which an integral part of
the instrument. Indeed it is critical to have a full system analysis
with the fringe tracker.  These servo-loop systems relax the
constraints on the VLTI interfaces by allowing for servo optical path
length differences and optimize the fiber injection of the input beams
to the required level. An internal optical switchyard allows the
operator to choose the best configuration of the VLTI co-phasing
scheme in order to perform phase bootstrapping for the longest
baseline on over-resolved objects. Three infrared science detectors
are implemented in the instrument, one for the Science Instrument, one
for the fringe tracker, and one for the tip-tilt sensor. The
instrument features 3 cryogenic vessels.

An important part of the instrument is the control system which
includes several servo-loop controls and management of the observing
software. The science software manages both data processing and image
reconstruction since one of the products of VSI will be a
reconstructed image like for the millimeter-wave interferometers. The
instrument development includes a plan for assembly, integration and
tests in Europe and in Paranal.

An instrument preliminary analysis report discusses several important
issues such as the comparison between the integrated optics and bulk
optics solutions, the standard 4- and 6-telescope VLTI array for
imaging, the proposed implementation of M12 mirrors to achieve these
configurations with VSI4 and VSI6, implication of using an
heterogeneous array and analysis of the thermal background.


The needs for future VLTI infrastructure can be summarized in an
increasing order of completeness as:
\begin{itemize}
\item {Interferometry Supervisor Software (ISS) upgrade}: upgrade from
  4-telescope version to a 6-telescope version allows VSI to use 6
  telescopes of the existing infrastructure for science cases which
  require imaging on a short timescale.
\item {AT5 and AT6:} 2 additional ATs allow the
  VLTI to use VSI in an efficient way without fast reconfiguration of
  the array.
\end{itemize} 
On a longer term, 8T combination at the VLTI could be foreseen but
this is not a VSI priority. In any case, it would require: 
\begin{itemize}
\item {DL7 and DL8:} 2 additional delay lines allow even
  without AT5 and AT6 to use all telescopes on the VLTI (4ATs+4UTs)
  and would be useful for complex imaging of rapidly changing
  sources. 
\item {AT7 and AT8:} could be implemented if DL7 and DL8 are
  procured. Then, the 8T VLTI capability could be exploited only with
  the ATs. 
\end{itemize}

The total cost of VSI for its 4-telescope version has been estimated
to about $4000$\,kEuros for hardware and a manpower of about 90 FTEs
over 4 years before the commissioning begins.The VSI6 version would
cost only $400$\,kEuros and 6 FTEs in addition to the VSI4 version.

\section{Conclusion}

The VLTI Spectro-Imager is an instrument whose science objectives is
to image astrophysical sources at the milli-arcsecond scale with both
high angular and high spectral resolutions in the near-infrared.
The image fidelity requirements drive a 6-beam design which 
makes use of the full potential of the VLTI site (including the use of
4 or 6 telescopes, use of PRIMA,...). VSI will deliver spectro-images
as final data product like in the ALMA scheme.

The design of the instrument is relatively simple and self-consistent
with its own fringe tracker in addition to other service modules like
tip-tilt, ADC,... The main science instrument is based on integrated
optics technology which has been shown to be as performant as a
solution bulk optics and has been sky validated. The fringe tracking
is an integral part of the imaging strategy.

In conclusion, the design of VSI is the simplest one that maximizes
science return for a large number of astrophysical domains interested
to phenomena at the milli-arcsecond scale.

\acknowledgments     

The VSI phase A study has benefited from a contract
by ESO, and support from JRA4 of OPTICON and from CNRS/INSU.
Amorim, Cabral, Garcia, Lima, Rebord\~ao,  was supported in part by the
Funda\,c\~ao para a Ci\^encia e a Tecnologia through project
PTDC/CTE-AST/68915/2006 from POCI, with funds from the European
programme FEDER. 


\bibliography{vsi-spie}   
\bibliographystyle{spiebib}   

\end{document}